\documentclass{iucrjournals}

\usepackage{amsmath}
\usepackage{amssymb}

\newcommand{\mytextin}[1]{\mbox{\scriptsize{#1}}}
\newcommand{\red}[1]{\textcolor{black}{#1}}
\newcommand{\blue}[1]{\textcolor{black}{#1}}

\title{Superstability of micrometer jets surrounded by a polymeric shell}

\author[a]{A. Rubio\IUCrEmaillink{arubiorg@unex.es}}%\IUCrOrcidlink{xxxx-xxxx-xxxx-xxxx}\IUCrAufn{Unique note.}}
\author[a]{J. M. Montanero\IUCrCemaillink{jmm@unex.es}}%\IUCrOrcidlink{xxxx-xxxx-xxxx-xxxx}\IUCrAufn{Shared note.}}
\author[b,c]{M. Vakili\IUCrEmaillink{mohammad.vakili@xfel.eu}}%\IUCrOrcidlink{0000-0001-9776-0177}\IUCrAufn{Unique note.}}
\author[b]{F. H. M. Koua\IUCrEmaillink{faisal.koua@xfel.eu}}%\IUCrOrcidlink{xxxx-xxxx-xxxx-xxxx}\IUCrAufn{Unique note.}}
\author[c,e]{S. Bajt\IUCrEmaillink{sasa.bajt@desy.de}}%\IUCrOrcidlink{xxxx-xxxx-xxxx-xxxx}\IUCrAufn{Unique note.}}
\author[c,d,e]{H.~N.~Chapman\IUCrEmaillink{henry.chapman@desy.de}}%\IUCrOrcidlink{xxxx-xxxx-xxxx-xxxx}\IUCrAufn{Unique note.}}
\author[f,g]{A. M. Ga\~n\'an-Calvo\IUCrEmaillink{amgc@us.es}}%\IUCrOrcidlink{xxxx-xxxx-xxxx-xxxx}}

\affil[a]{Depto.\ de Ingenier\'{\i}a Mec\'anica, Energ\'etica y de los Materiales and Instituto de Computaci\'on Cient\'{\i}fica Avanzada (ICCAEx),
Universidad de Extremadura, E-06006 Badajoz, Spain}
\affil[b]{European XFEL, Holzkoppel 4, 22869 Schenefeld, Germany}
\affil[c]{Center for Free-Electron Laser Science CFEL, Deutsches Elektronen-Synchrotron DESY, Notkestr. 85, 22607 Hamburg, Germany}
\affil[d]{Department of Physics, University of Hamburg, Luruper Chaussee 149, 22761 Hamburg, Germany}
\affil[e]{The Hamburg Centre for Ultrafast Imaging, Luruper Chaussee 149, 22761 Hamburg, Germany}
\affil[f]{Departamento de Ingenier\'{\i}a Aeroespacial y Mec\'anica de Fluidos, Universidad de Sevilla, E-41092 Sevilla, Spain}
\affil[g]{ENGREEN, Laboratory of Engineering for Energy and Environmental Sustainability, Universidad de Sevilla, 41092 Sevilla, Spain.}

\begin{document} 
\maketitle 

\begin{synopsis}
Superstable micrometer compound jets are produced for Serial Femtosecond X-ray Crystallography
\end{synopsis}

\begin{abstract}
We have produced superstable compound liquid microjets with a three-dimensional printed coaxial flow-focusing injector. The aqueous jet core is surrounded by a shell, a few hundred nanometers in thickness, of a low-concentration aqueous solution of a low-molecular-weight polymer. Due to the stabilizing effect of the polymeric shell, the minimum liquid flow rate leading to stable flow-focusing is decreased by one order of magnitude, resulting in much thinner and longer jets. Possible applications of this technique for Serial Femtosecond X-ray Crystallography are discussed.
\end{abstract}

%\keywordsSerial{flow focusing;Femtosecond X-ray Crystallography}

\section{Introduction}
\label{sec1}

% XFELS
X-ray free-electron lasers produce highly energetic and coherent pulses of X-rays with femtosecond duration. By providing a probe with a wavelength that can access atomic scales with a pulse commensurate with the time scales of atomic motions, these machines have opened new ways to study the structure and dynamics of condensed matter, liquids, and gases, covering fields ranging from planetary science, atomic physics, and structural biology~\cite{Bostedt:2016}. When the X-ray pulses are tightly focused, the resulting extreme intensity vaporizes material, so many experiments are carried out by rapidly replacing or refreshing the exposed sample prior to the next pulse. The first macromolecular crystallography experiments conducted at a hard X-ray free-electron laser (XFEL) faced the challenge of introducing hydrated protein crystals to the X-ray beam, in a vacuum environment, to be exposed and refreshed at a rate of 120 pulses per second \cite{Chapman:2011}. This was achieved using a liquid jet of a slurry of nanocrystals, formed by gas focusing \cite{G98a} using a gas dynamic virtual nozzle \cite{DWSWSSD08}. Since those first experiments, liquid jets have been used for many serial femtosecond crystallography (SFX) experiments at all XFEL facilities around the world, and developments of nozzles and microfluidic mixers \cite{Knoska:2020,Doppler2023} have enabled mix-and-inject crystallography \cite{Schmidt:2013} and pump–probe time-resolved crystallography of photoactive proteins \cite{Branden:2021}. The same instrumentation is also used for solution scattering \cite{Arnlund:2014,Blanchet:2023} and fiber diffraction \cite{Popp:2017},  and similar jets are used in spectroscopy experiments of liquids \cite{Lemke:2017}. The high speeds, of more than 50~m/s that micrometer-diameter gas-focused jets can reach \cite{Knoska:2020}, enabled the adoption of serial crystallography with the megahertz pulse trains of the European XFEL \cite{Wiedorn:2018}.

Many innovations and variations of liquid ejection have been developed for the priorities of specific experiments, such as acoustic injection to reduce sample consumption at e.g. 120 Hz repetition rate \cite{Roessler:2016}, and sheet jets for reduced scattering background and to provide a constant interaction volume with the X-ray beam for spectroscopy measurements \cite{Konold2023}. However, the basic cylindrical jet configuration offers multiple advantages such as robustness, ease of use, consistency, high sample velocities (for high repetition rates of X-ray pulses), and compatibility with a wide range of sample types. Designs for cylindrical micro-jets have therefore continued to evolve towards \blue{a more stable and efficient sample delivery} as well as support of measurements over an increased range of time spans. A strategy to achieve this is to separate the dual functions of the liquid as a medium to carry crystals and to form a stable jet, by using a concentric flow of two liquids \cite{G98a}. This was realized by using a double-flow focusing nozzle (DFFN) to flow an outer sheath of liquid of low surface tension such as ethanol over the central buffer containing the crystals, all focused by gas \cite{Oberthuer:2017}. The flow of the sample-containing liquid can be reduced arbitrarily without affecting the jet, which is formed by the sheath liquid. 

Here, we extend the concept of utilizing two fluids to achieve even higher control and stability of the flowing jet, and to reach a wider span of time delays by extending the jet length. Until now, most of the liquids used in sample delivery exhibit quasi-Newtonian mechanical properties, which necessitates increasing the viscosity of the buffer to lengthen the jet and enhance its stability. It has been recently shown that diluted solutions of specific polymers can significantly extend the jet length while reducing the liquid flow rate by an order of magnitude \cite{RGVMC22,RVGM22}. These polymers can adhere to the sample requirements. This method presents a new opportunity to improve the conventional approaches in liquid sample handling formulations. 

% This study
In this study, we \blue{apply} a non-invasive method in which a pre-existing buffer formulation is maintained within the core of the jet. The core is enveloped by a shell (film) of a miscible low-concentration water solution containing a compatible polymer. We aim to determine whether the increased jet length observed in simple jets occurs in a compound or armored jet, akin to a continuously flowing sausage. This setup required implementing a concentric flow focusing technique \cite{G98a} utilized in previous research \cite{Oberthuer:2017,Vakili:2022}. Our findings confirm our expectations and reveal unexpected benefits that significantly enhance the importance of the involved nonlinear fluid-dynamics phenomena, meriting further detailed examination.

\section{Methods}
\label{sec2}

% Setup
The gaseous flow-focusing ejector employed in this study was printed using the Nanoscribe Photonic Professional GT2 with the Dip-in Laser Lithography (DiLL) configuration, a 25$\times$ objective lens, and the IP-S photoresist. \blue{The typical slicing and hatching distances were 1 $\mu$m and 0.5 $\mu$m, respectively. These distances are much smaller than the lengths characterizing the nozzle. Therefore, one does not expect the nozzle fabrication reproducibility to affect the results.}

\begin{figure}[!tbp]
\begin{center}
\resizebox{0.7\columnwidth}{!}{\includegraphics{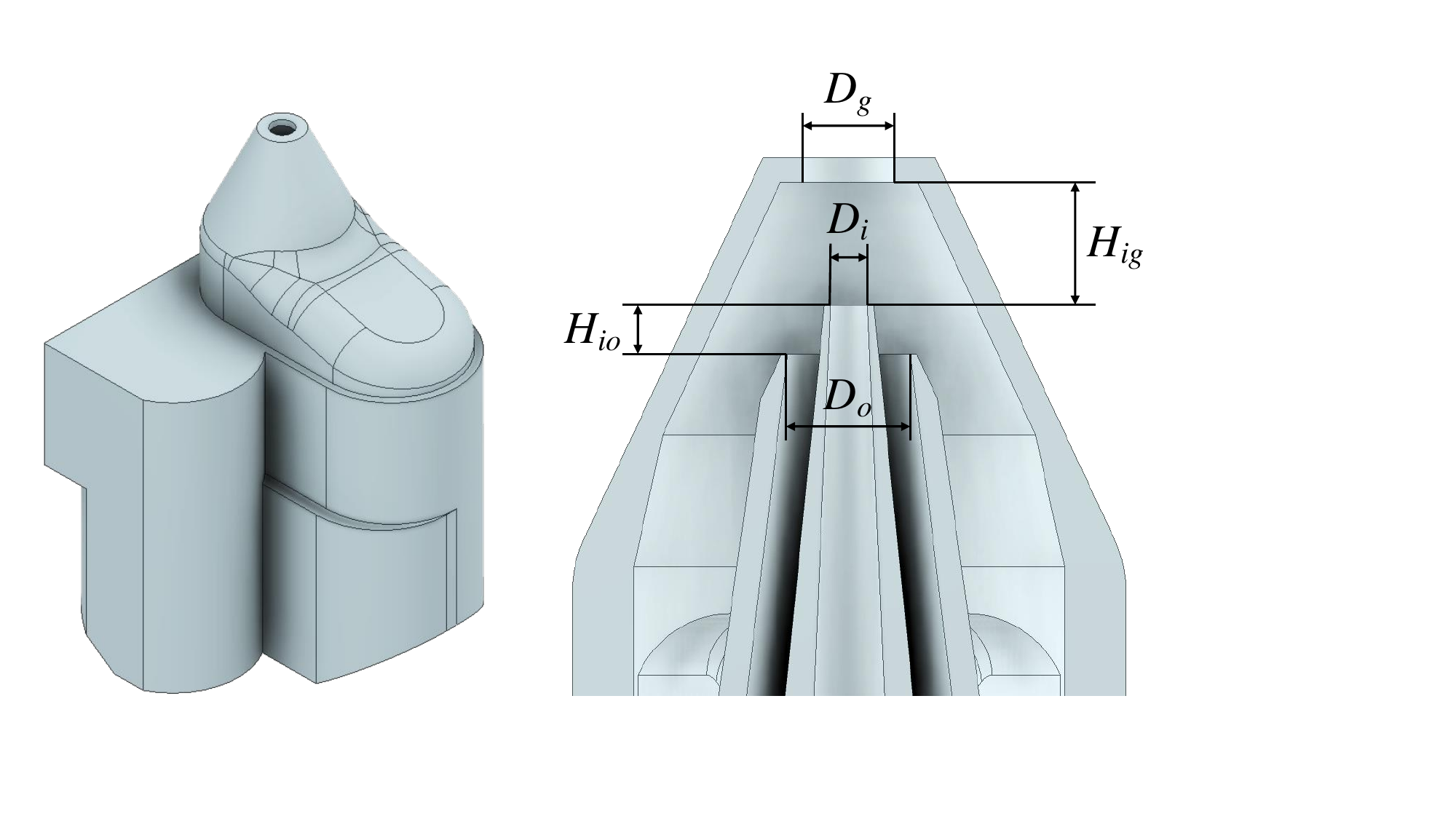}}
\end{center}
\caption{CAD screenshot of the nozzle with indications of geometric design parameters.}
\label{geom}
\end{figure}

The ejector consisted of a concentric annular duct coaxially placed inside a converging nozzle (Fig.\ \ref{geom}). The focused liquid was injected across the inner capillary at a constant flow rate $Q_i$ with a syringe pump (LEGATO210, KD-SCIENTIFIC). The fluid shell was injected through the outer capillary at a constant flow rate $Q_o$ with another identical syringe pump. The gaseous stream was helium. The helium mass flow rate was measured with a mass flow meter (FLOW-BUS, BRONKHORST). We mounted the ejector onto the cap of a discharge glass cell and established a pressure of 25 mbar inside the cell using a suction pump.

% Images
We used a high-speed video camera (Fastcam Mini UX50) to take digital microscopy images of the jet at $4000$ fps. The camera was equipped with optical lenses ({\sc Optem Zoom 70XL}) and a microscope objective ({\sc Optem} 10X). The images consisted of 974$\times$768 pixels. The magnification was 4.5$\times$, which resulted in 0.222 $\mu$m/pixel. We illuminated the jet with a laser (SI-LUX 640, {\sc Specialized Imaging}) synchronized with the camera, which reduced the effective exposure time down to 150 ns. To examine the jet behavior at different jet stations, the camera was displaced vertically using a triaxial translation stage with its vertical axis motorized ({\sc Thorlabs} Z825B) and controlled by the computer. The images were analyzed to measure the jet diameter with pixel resolution at different jet stations. For some representative cases, we repeated the experiment several times. We concluded that the experimental uncertainty in the jet diameter measurement was around one-pixel size.

% Liquids
%We conducted experiments with water ($\rho=998$ kg/m$^3$ and $\eta=1$ mPa$\cdot$s) and a typical buffer fluid used in protein crystallography measurements as core fluid. \blue{For the sheath liquid, a Newtonian aqueous polyethylene oxide (PEO) solution was used. At 100\% solids, the utilized PEO (PEO100K) has a of density $\rho=1.13$ g/cm$^3$ and viscosity $\eta=17.5$ mPa$\cdot$s.} 

% Liquids
In our flow-focusing experiments, we produced (i) single jets of water ($\rho=998$ kg/m$^3$ and $\mu=1$ mPa$\cdot$s) and an aqueous buffer fluid ($\mu=4$ mPa$\cdot$s) as well as (ii) compound jets of those liquids surrounded by either those fluids or the polymeric shell. \blue{Details of the polymeric shell are given in Sec.\ \ref{sec3}}. The buffer fluid is a typical carrier medium used in protein crystallography.

% Jet diameter and length
In each experimental run, we fixed the inner liquid flow rate $Q_i$ and progressively reduced the outer flow rate $Q_o$ until the stable ejection ceased. We considered a stable ejection if there was no interruption in the jet for more than 40 s. We measured the jet length $L_j$ and diameter $d_j$ at the stability limit. The jet length $L_j$ was determined as the distance from the ejector at which the capillary varicose instability arose. The jet diameter $d_j$ was measured at that distance from the ejector. The jet inner (core) diameter $d_i$ was calculated as $d_i=d_j \sqrt{Q_i/Q_t}$, where $Q_t=Q_i+Q_o$ is the total liquid flow rate. Here, we have assumed that the core and shell moved at the same velocity, which is an accurate approximation given the small size of the laminar entrance length $L_e/d_j\simeq 0.0575\, \text{Re}\sim 10$, where Re$=\rho d_j v_j/\mu\sim 10^2$ is the jet Reynolds number. The thickness of the shell was calculated as $t=(d_j-d_i)/2$. 

% Boas and Whipping
We observed the jet at each camera position over several seconds to determine the position at which the Beads-on-a-String (BOAS) \cite{GYPS69} and whipping \cite{MG20} instabilities appeared. We considered that the BOAS instability was present if the diameter increased by more than 10\% of its average value in that jet station. We considered that the jet suffered from whipping instability if the amplitude of its lateral oscillations exceeded the threshold $0.1\, d_j$. The appearance of beads was observed occasionally.

% Freezing
\blue{Numerical simulations \cite{RRCHGM21} have shown that the jet temperature does not significantly change around the nozzle despite the low temperature of the gaseous jet. This occurs because of the small residence time of a fluid element within this region.}
 
\section{Optimization Study}
\label{sec3}

% Geometry
A systematic study preceded the analysis presented in this work to determine the optimal conditions for the jet ejection. Eight nozzles were designed to maximize the stabilizing effect of the polymeric shell. Table \ref{nozzles} shows the characteristic lengths of those nozzles (see Fig.\ \ref{geom}). \blue{These lengths were changed according to the following observations: $H_{ig}$ was reduced with respect to values typically used with Newtonian liquids because the tapering meniscus retracts (moves away from the discharge orifice) due to the viscoelastic sheath. $H_{io}$ was reduced to prevent the sheath from touching the inner nozzle. $D_i$ and $D_o$ were increased in N9 to avoid nozzle clogging during crystal jetting.} 

\begin{table}
\begin{tabular}{cccccc}
Nozzle & $D_i$ ($\mu$m) & $D_o$ ($\mu$m) & $D_g$ ($\mu$m)& $H_{ig}$ ($\mu$m) & $H_{io}$ ($\mu$m)\\
\hline
N1 & 30 & 100 & 75 & 100 & 40 \\
\hline
N2 & 30 & 100 & 75 & 90 & 50 \\
\hline
N3 & 20 & 100 & 75 & 90 & 50 \\
\hline
N4 & 30 & 100 & 75 & 130 & 10 \\
\hline
N5 & 30 & 100 & 75 & 140 & 0 \\
\hline
N6 & 30 & 75 & 75 & 100 & 40 \\ 
\hline
N7 & 30 & 100 & 75 & 80 & 10 \\
\hline
N8 & 30 & 100 & 75 & 80 & 30 \\
\hline
N9 & 75 & 120 & 60 & 130 & 10\\
\hline
\end{tabular}
\caption{Characteristic lengths of the nozzles used in the experiments}
\label{nozzles}
\end{table}

We conducted experiments with all the nozzles. In the experiments with water, nozzles N1, N3, and N5 enhanced the whipping of the fluid meniscus (absolute whipping \cite{AFMG12}), nozzle N2 produced jets with larger diameters, and nozzles N7 and N8 led to an increase in the minimum total flow rate. For these reasons, we selected nozzle N4 to focus water jets. In the case of the buffer fluid, nozzle N2 produced jets with larger diameters, nozzles N3--N5 enhanced the meniscus and jet whipping, and nozzles N6-N8 led to an increase in the minimum flow rate. For these reasons, we selected nozzle N1 to produce the buffer fluid jets. \blue{It must be noted that there is space for improving the nozzle design. The influence of parameters, such as the nozzle converging rate, was not analyzed}. 

% PEO 100K c=1%
\citeasnoun{RGVMC22} examined the effect of polyethylene oxide (PEO) dissolved in aqueous solutions on the stability of a single-jet produced with gaseous flow focusing. They achieved the best results with the molecular weight $M_w=100\times 10^3$ g/mol (PEO100K) at the concentration $c=1\%$ (wt). \citeasnoun{RVGM22} also considered polyvinylpyrrolidone (PVP) with $M_w=360\times 10^3$ g/mol (PVP360K), but it led to a lower reduction of the minimum flow rate due to the pull-out instability \cite{WHBMB12}. These previous results suggest that PEO100K at $c=1\%$ (wt) may produce the best results in the coaxial configuration studied in the present work. To confirm this hypothesis, we conducted experiments with PEO (Sigma-Aldrich, product no. 181986) at concentrations between 0.3\% (wt) and 1.5\% (wt). The best results were obtained for $c=1\%$ (wt), as occurred with the single-jet configuration \cite{RGVMC22,RVGM22}. Smaller concentrations did not produce significant stabilizing effects, while larger concentrations led to pull-out instability \cite{WHBMB12}. PEO100K at $c=1\%$ (wt) can be regarded as a low-viscosity Boger (shear-thinning free) fluid \cite{RGVMC22}. Its shear viscosity is $\mu=2.33$ mPa$\cdot$s while its extensional relaxation time is $\lambda_r=26.2$ $\mu$s \cite{RGVMC22}.

\begin{figure}[!tbp]
\begin{center}
\resizebox{0.45\columnwidth}{!}{\includegraphics{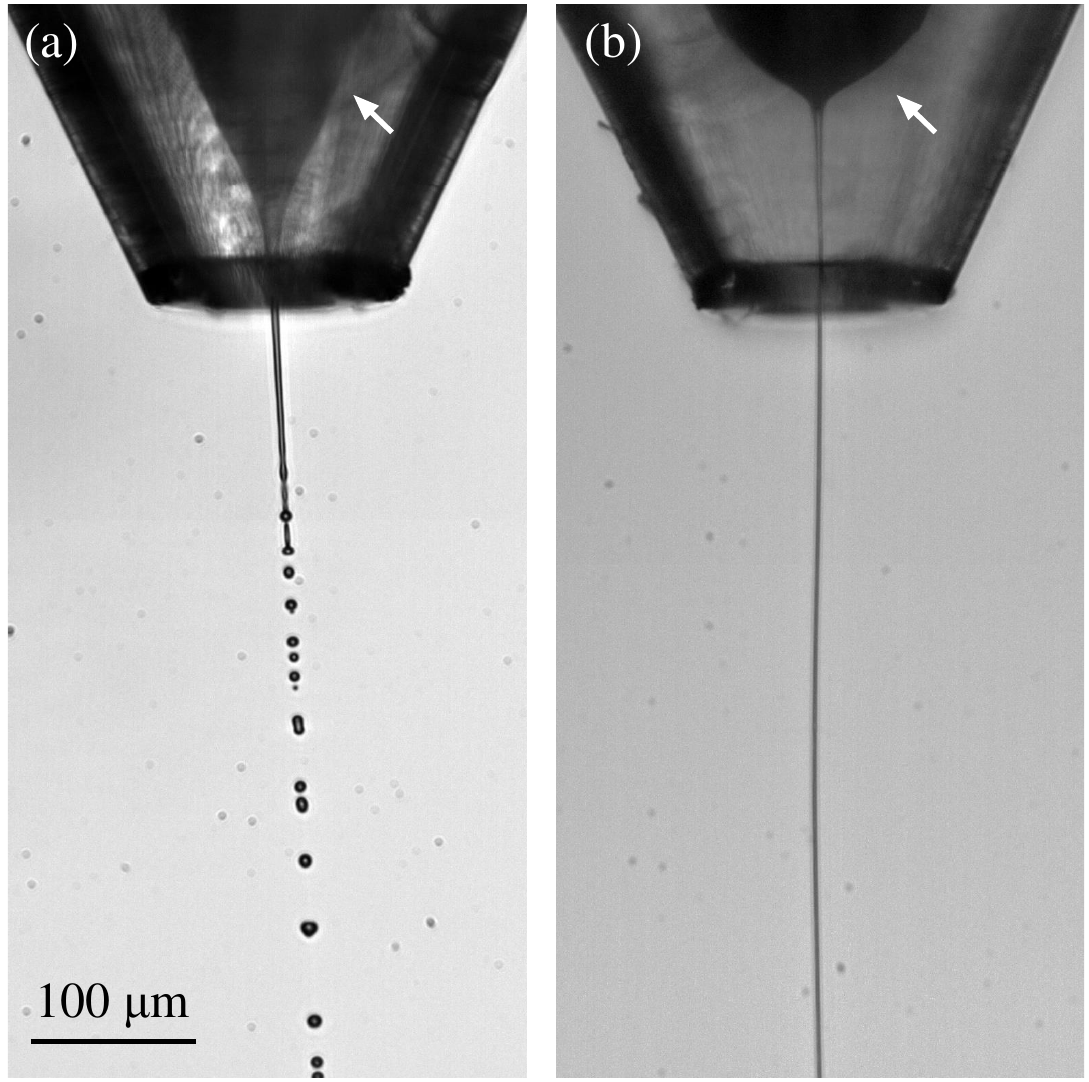}}
\end{center}
\caption{Images of the water tapering meniscus (marked with an arrow) and emitted jet surrounded by water (a) and the viscoelastic shell (b). The images were obtained for $Q_i=1$ $\mu$l/min and $Q_o=13$ $\mu$l/min (a) and $Q_i=1.5$ $\mu$l/min and $Q_o=1.75$ $\mu$l/min (b). The mass flow rate was $m'_g=18.5$ mg/min in the two cases. \blue{The images show experiments conducted with the nozzle N4}.}
\label{shrink}
\end{figure}

% Mass flow rate
Experiments were conducted with four gaseous mass flow rates in the range of $10-25$ mg/min. We selected $m'_g=18.5$ mg/min and 23.8 mg/min for water and the buffer liquid, respectively. Increasing (decreasing) the mass flow rate above (below) these values enhanced the whipping (BOAS) instability. Once the rest of the variables have been fixed, the only two control parameters are the inner $Q_i$ and outer $Q_o$ flow rates. For the sake of illustration, Fig.\ \ref{shrink} shows images of the water jet surrounded by water (a) and the viscoelastic shell (b).

\section{Results and Discussion}
\label{sec4}

\begin{figure}[!tbp]
\begin{center}
\resizebox{0.6\columnwidth}{!}{\includegraphics{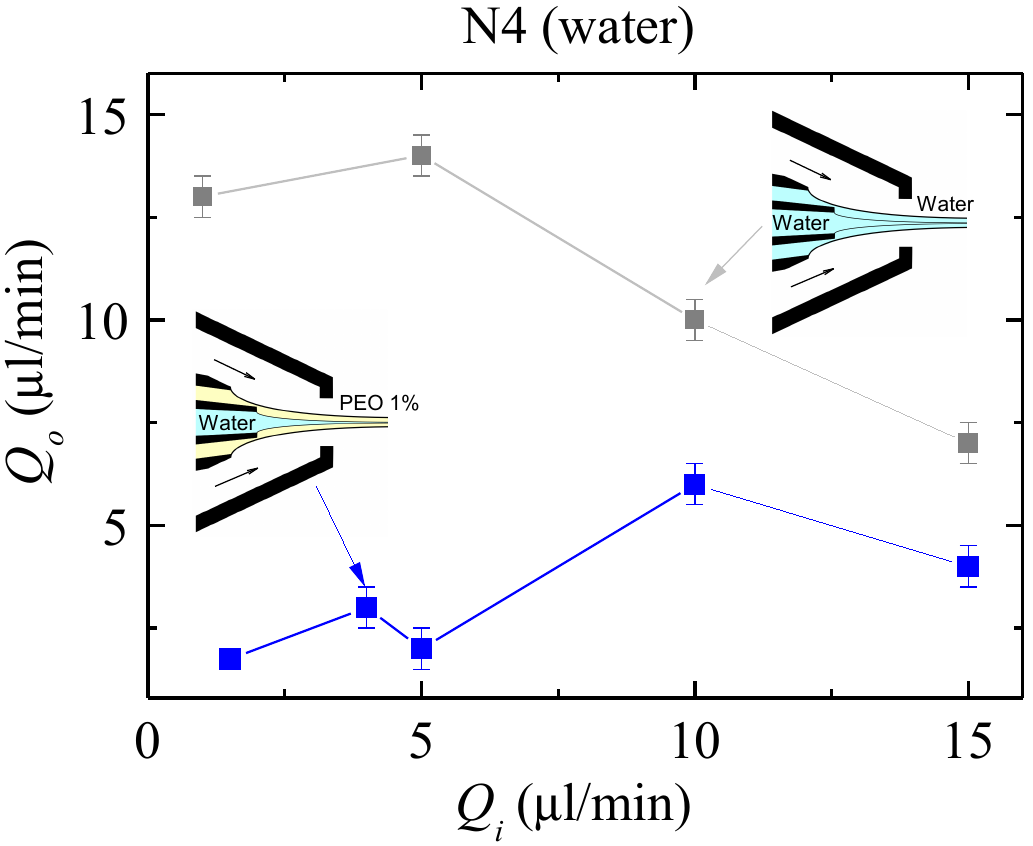}}
\end{center}
\caption{Outer flow rate $Q_o$ as a function of the inner flow rate $Q_i$ at the stability limit for the water jet surrounded by the water shell (gray squares) and the viscoelastic shell (blue squares).}
\label{water3}
\end{figure}

\begin{figure}[!tbp]
\begin{center}
\resizebox{0.6\columnwidth}{!}{\includegraphics{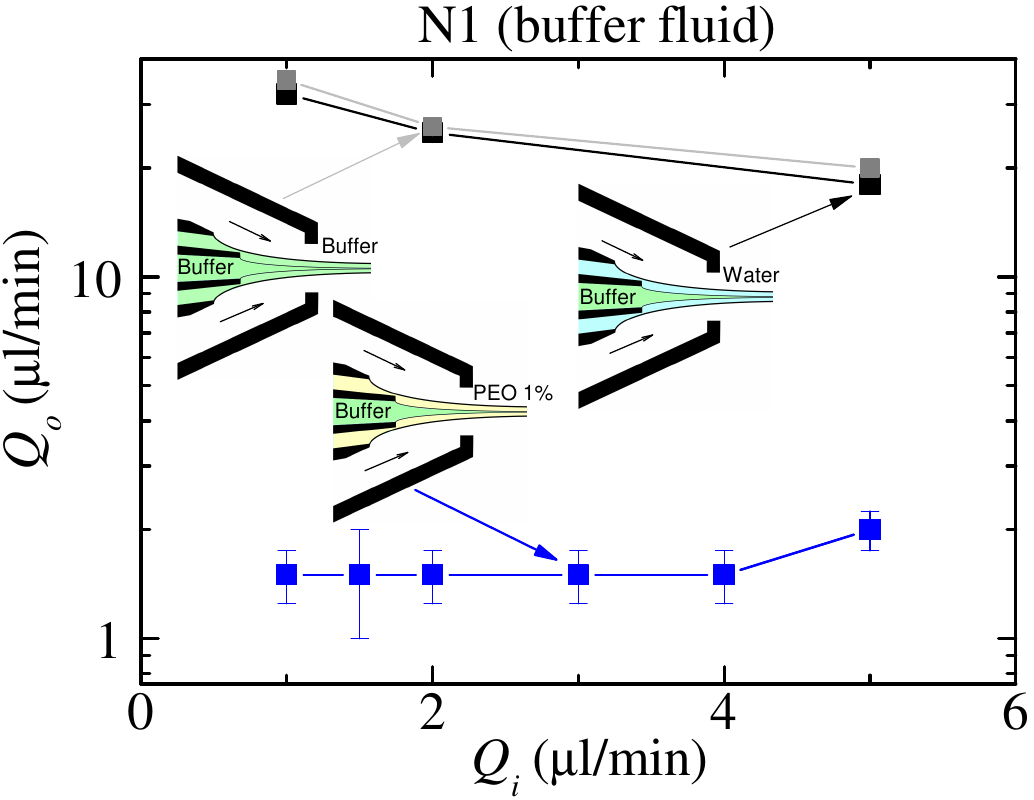}}
\end{center}
\caption{Outer flow rate $Q_o$ as a function of the inner flow rate $Q_i$ at the stability limit for the buffer fluid jet surrounded by the water shell (gray squares), the buffer fluid shell (black squares), and the viscoelastic shell (blue squares).}
\label{buffer3}
\end{figure}

% Minimum flow rate
We started our study by determining the stability limit for each fluid configuration. Figures \ref{water3} and \ref{buffer3} show the outer flow rate $Q_o$ below which the jet emission becomes unstable for a given inner flow rate $Q_i$. The figures show the strong stabilizing effect of the viscoelastic (polymeric) shell compared with that produced by water and the buffer fluid, especially when the inner liquid is the buffer fluid (Fig.\ \ref{buffer3}). In this case, the outer liquid flow rate can be decreased by one order of magnitude when the polymer is added, which leads to a large reduction of the minimum total flow rate $Q_t$. This effect has been previously observed in simple jets \cite{RGVMC22,RVGM22}, and can be explained by the coil-stretch transition \cite{D74} undergone by the polymers in the tapering meniscus. Despite the short nature of the polymer relaxation time, the large strain rate produced by the transonic gas stream manages to stretch the polymers for the selected parameter conditions. The elastic stress shrinks the meniscus (Fig.\ \ref{shrink}b) and stabilizes the flow, which reduces the minimum total flow rate for which the steady jetting regime can be obtained. As shown below, the polymeric shell produces this noticeable effect even though its thickness is one order of magnitude smaller than the jet diameter. 

Figure \ref{buffer3} compares the effects on the meniscus stability of the shells of water and the buffer fluid, two Newtonian liquids with very different viscosities. As observed, they produce similar effects. However, the viscoelastic shell produces a much more significant reduction of the outer flow rate, even though its zero-shear viscosity is very similar to that of water. This confirms that the meniscus stabilization must be attributed to viscoelasticity. 

\begin{figure}[!tbp]
\begin{center}
\resizebox{0.6\columnwidth}{!}{\includegraphics{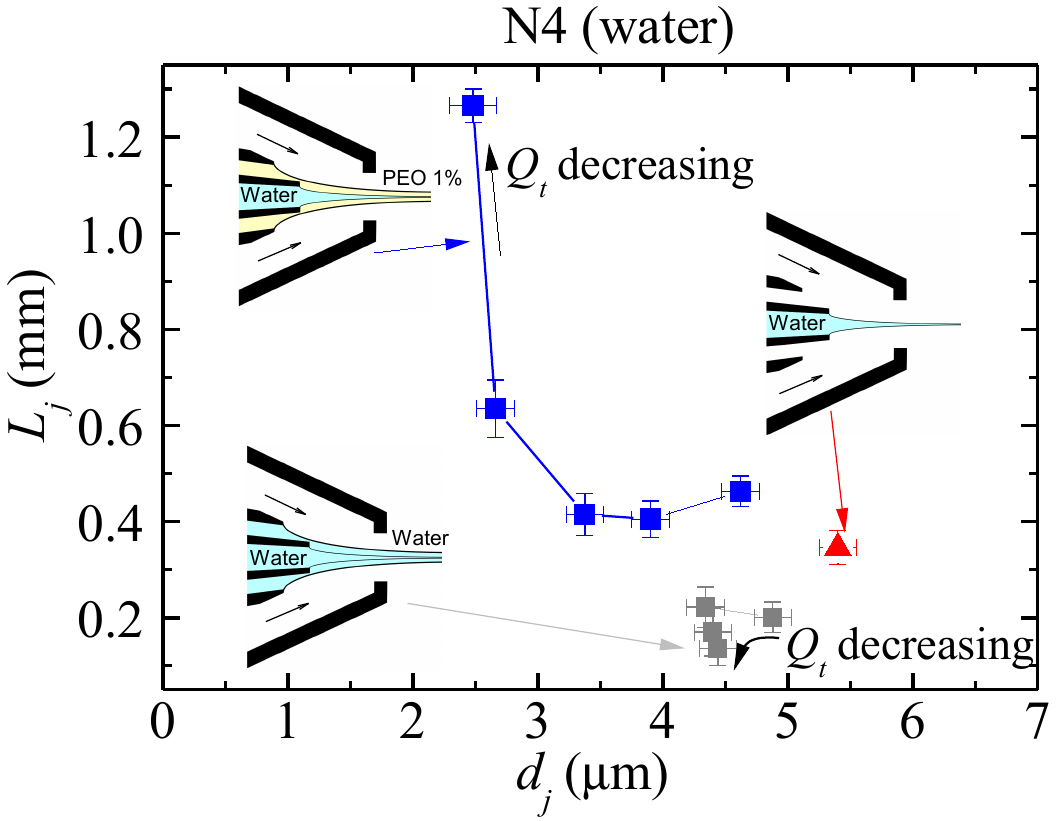}}
\end{center}
\caption{Jet length $L_j$ versus jet diameter $d_j$ at the stability limit for the water jet alone (red triangle) and the water jet surrounded by the water shell (gray squares) and the viscoelastic shell (blue squares).}
\label{water}
\end{figure}

\begin{figure}[!tbp]
\begin{center}
\resizebox{0.6\columnwidth}{!}{\includegraphics{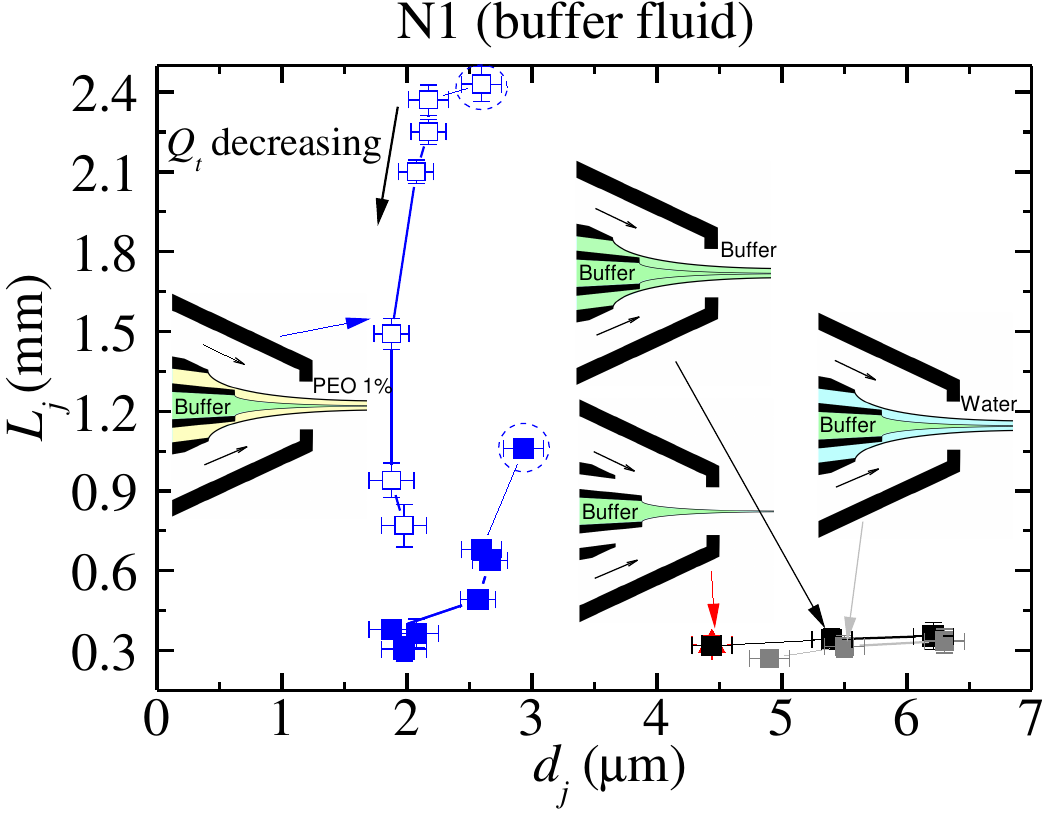}}
\end{center}
\caption{Jet length $L_j$ versus jet diameter $d_j$ at the stability limit for the buffer fluid jet alone (red triangle) and the buffer fluid jet surrounded by the buffer fluid shell (black symbols), the water shell (grey symbols), and the viscoelastic shell (blue squares). The open symbols correspond to the results without considering the jet whipping. \red{The points marked with a circle were obtained for $Q_i=5$ $\mu$l/min and the outer flow rate $Q_o=5$ $\mu$l/min, which is above its minimum value $Q_o=2$ $\mu$l/min.}}
\label{buffer}
\end{figure}

% Diameter and length
We aim to produce jets that are as thin and long as possible. Figures \ref{water} and \ref{buffer} show the noticeable positive effect of the thin viscoelastic shell for this purpose. Much thinner and longer jets were produced when the polymer was added to the outer liquid, even though the polymer's molecular weight and concentration were very small. \red{Figure \ref{buffer} also shows an experimental realization with an outer flow rate above its minimum value. Increasing the outer flow rate stabilizes the jet.} 

% Diameter
The polymeric shell allowed us to reduce the minimum total flow rate (Figs.\ \ref{water3} and \ref{buffer3}) and, therefore, the minimum jet diameter. Buffer fluid jets around 2 $\mu$m in diameter were produced with the polymeric shell, while $d_j\gtrsim 4.5$ $\mu$m in the absence of the outer viscoelastic liquid. A similar reduction in the minimum jet diameter was obtained in the water case.

% Length
Water jets with lengths hundreds of times the diameter were ejected due to the stabilizing effect of the viscoelastic shell. These jets suffered neither the BOAS nor the whipping instability. Buffer fluid jets with lengths hundreds of times the diameter were also steadily emitted. The jet core viscosity and ambient pressure triggered the whipping instability at distances from the ejector between 0.3 and 1 mm, approximately. It must be pointed out that this convective instability did not alter the meniscus flow stability and can be partially suppressed by reducing the ambient pressure. If the whipping instability is obviated, micrometer jets with lengths larger than 2 mm were steadily emitted owing to the stabilizing effect of the polymeric shell.

% Flight time
The residence time of a fluid particle before reaching the breakup jet station can be estimated as $t_r\simeq L_j/v_j$, where $v_j\simeq 4Q_t/(\pi d_j^2)$ is the jet velocity. Consider, for instance, the buffer fluid jet with $d_j=2.1$ $\mu$m and $L_j=2.1$ mm (Fig.\ \ref{buffer}) ejected with $Q_t=4.5$ $\mu$l/min (Fig.\ \ref{buffer3}). In this case, $t_r\simeq 95$ $\mu$s, which is much larger than the polymer (extensional) relaxation time $\lambda_r=26.2$ $\mu$s. We hypothesize that the strong converging flow in the meniscus causes some entanglement or self-assembly of the stretched polymers carried by the shell, which hinders the polymer (extensional viscosity) relaxation and delays the capillary instability.

\begin{figure}[!tbp]
\begin{center}
\resizebox{0.6\columnwidth}{!}{\includegraphics{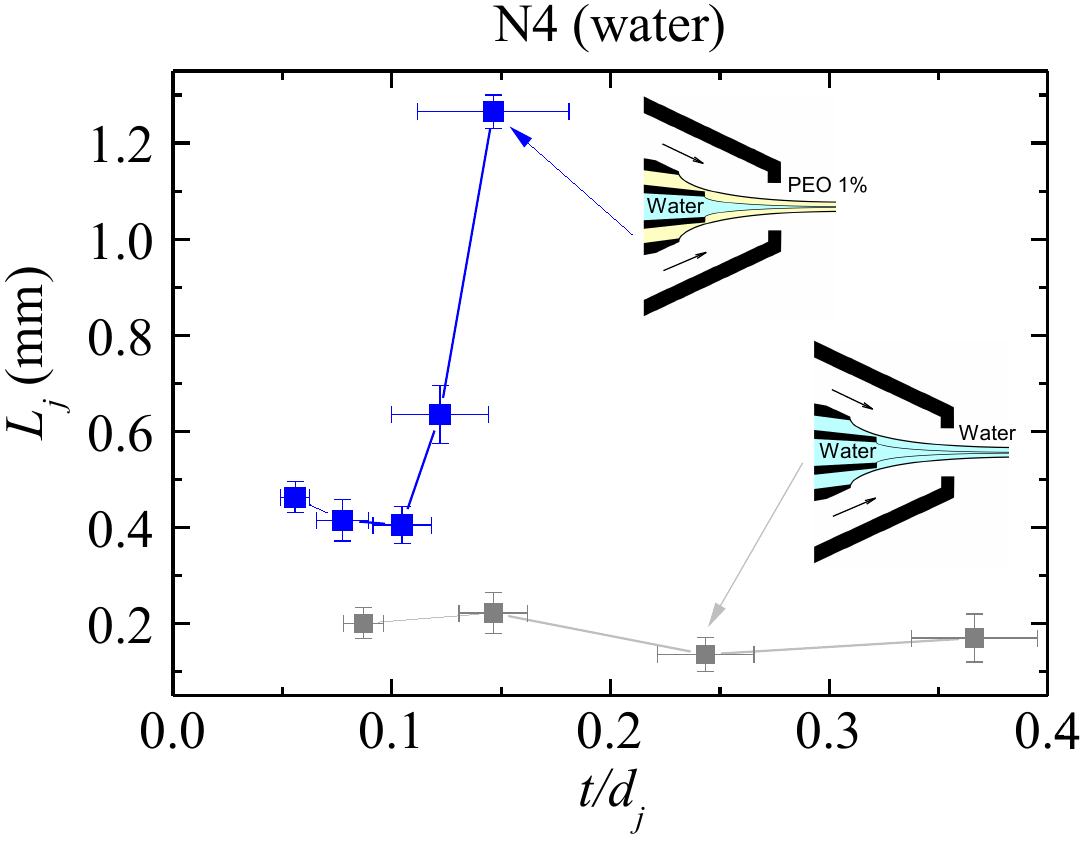}}
\end{center}
\caption{Jet length $L_j$ versus jet ratio $t/d_j$ at the stability limit for the water jet surrounded by the water shell (grey squares) and the viscoelastic shell (blue squares).}
\label{water2}
\end{figure}

\begin{figure}[!tbp]
\begin{center}
\resizebox{0.6\columnwidth}{!}{\includegraphics{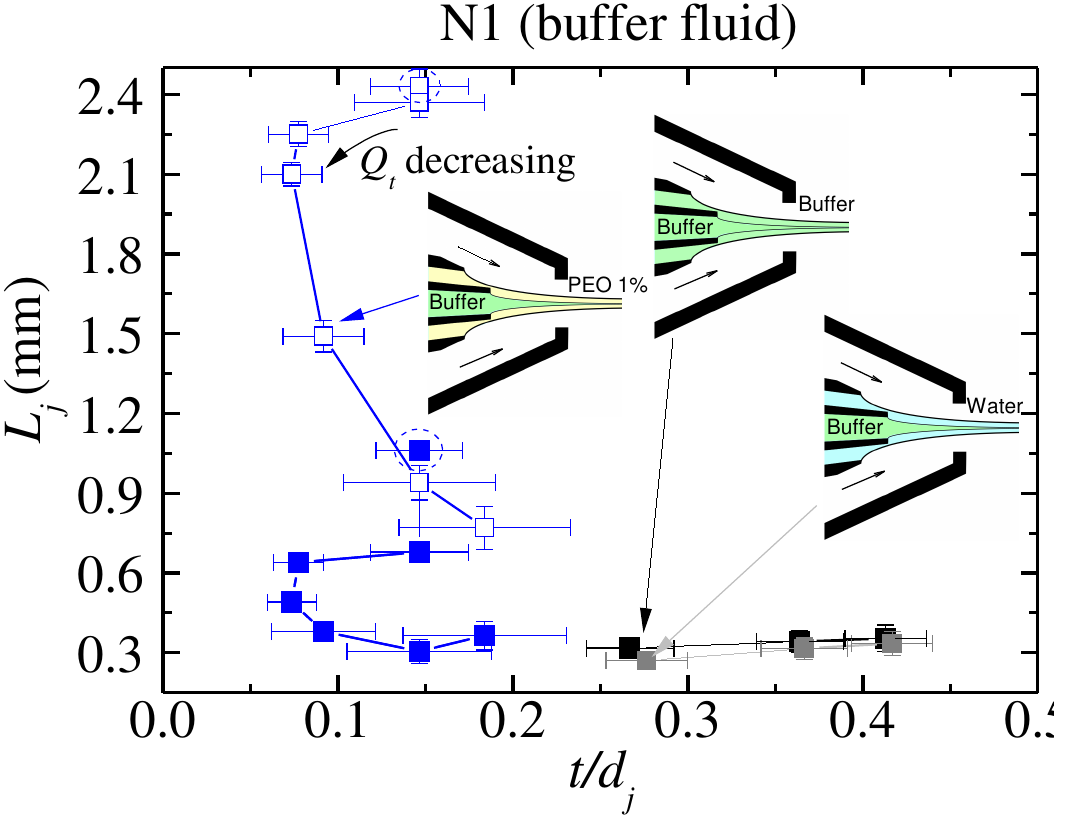}}
\end{center}
\caption{Jet length $L_j$ versus jet the ratio $t/d_j$ at the stability limit for the buffer fluid jet surrounded by the buffer fluid shell (black symbols), the water shell (grey symbols), and the viscoelastic shell (blue squares). The open symbols correspond to the results without considering the jet whipping. \red{The points marked with a circle were obtained for $Q_i=5$ $\mu$l/min and the outer flow rate $Q_o=5$ $\mu$l/min, which is above its minimum value $Q_o=2$ $\mu$l/min.}}
\label{buffer2}
\end{figure}

% Thickness
Finally, we plot the results as a function of the shell relative thickness $t/d_j$ in Figs.\ \ref{water2} and \ref{buffer2}. Interestingly, the viscoelastic shell produces a strong stabilizing effect even though its thickness $t$ is much smaller than the jet diameter $d_j$. In fact, the relative thickness of the viscoelastic shells is smaller than that of the Newtonian ones in the buffer fluid case.

\begin{figure}[!tbp]
\begin{center}
\resizebox{0.7\columnwidth}{!}{\includegraphics{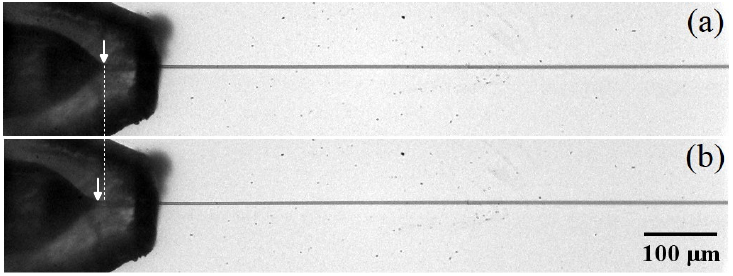}}
\end{center}
\caption{Images of (a) water and (b) buffer with protein microcrystal jets surrounded by the viscoelastic shell. The images were obtained with N9 for \blue{$Q_i=5$ $\mu$l/min, $Q_o=12$ $\mu$l/min and $m'_g=15.3$ mg/min.} The crystal concentration was adjusted to be $c_{\mytextin{crys}}\simeq 15-20\%$ (v/v) after settlement. The arrows approximately indicate the tip of the tapering meniscus.}
\label{crystals}
\end{figure}

% Application
Our study has shown that using diluted solutions of specific polymers significantly increases the jet length while reducing the liquid flow rate by one order of magnitude. To illustrate the capabilities of this method, Fig.\ \ref{crystals} shows images of the jet emitted with nozzle N9. This design variation provides a larger $D_i$ to facilitate microcrystal compatibility (see also the video in the Supplemental Material). It led to remarkable results with both water and the buffer fluid. Crystals of the protein C-phycocyanin (10$\times$3$\times$2 $\mu$m ($\pm 25\%$ all axes)) were mixed in an aqueous buffer \blue{(75 mM HEPES (pH 7.0), 20 mM MgCl$_2$ and $25\%$ (w/v) PEG3350)} at an adjusted concentration of $c_{\mytextin{crys}}\simeq 15-20\%$ (v/v) after settlement. With a total flow rate of \blue{12 $\mu$l/min, and a gas flow rate of $m'_g=15.3$ mg/min, the jet diameter was ca. 3 $\mu$m, the jet velocity about 41.2 m/s, and the length larger than 800 $\mu$m for water and the buffer jet. The inner liquid flow rate was 5 $\mu$l/min in the two experiments.} The difference between the jet cores hardly affected the shape of the tapering meniscus and jet (the buffer fluid meniscus was slightly shorter). The amplitude of the jet lateral oscillation (whipping) was comparable to the jet diameter for \blue{both water and buffer jet (see supplemental videos)}. With a constant inner flow rate, stable jets with a sheath flow as low as $Q_o=2$ $\mu$l/min could be obtained with N9.

% Mo's comments
In today's SFX experiments that utilize liquid jets in vacuum, the length of the jet is usually less than 500~$\mu$m and even shorter (about 200-300 $\mu$m) when high liquid speeds are needed for megahertz X-ray pulse repetition rates. In such experiments, the nozzles are positioned so that the X-rays (with a focus size of a few micrometers) intersect the uninterrupted liquid column, flowing perpendicular to the X-ray beam, at the largest possible distance from the nozzle tip while avoiding the droplet regime. Time-resolved SFX experiments using pump–probe schemes utilize femtosecond-duration optical laser pulses focused to spot sizes of between 20 and 100 $\mu$m to trigger structural changes of the jetted sample prior to the arrival of the X-ray pulse. By varying the delay between the two pulses, structural dynamics can be mapped at sub-nanosecond times and longer \cite{Pandey:2020}. However, depending on the speed of the jet, delays longer than about 1~$\mu$s require the optical pump to be focused on the jet upstream of the interaction region. Pumping upstream or even inside the nozzle can be problematic because of liquid recirculation in the converging part of the nozzle, which disperses the arrival times of photoexcited crystals at the X-ray interaction point, spoiling the time resolution. Furthermore, the polymeric 3D printed material is often degraded by the laser pulses, especially in vacuum, where there is no heat dissipation. Therefore, an ultralong and stable jet can increase the upper limit of time delays that can be accessed with high precision. For instance, a \blue{1.5 mm long jet would allow the capture of time delays of 25 $\mu$s (assuming a laser-to-X-ray spacing of 1.2 mm and a jet velocity of 45 m/s as usually employed with megahertz X-ray pulses).} Photoactivation time delays longer than that (few hundred of $\mu$s-milliseconds) can be realized using synchrotron radiation with a range of other sample delivery methods  \cite{SYPM22}. 

Viscoelastic jets appear to be very promising for SFX experiments due to their length and small diameter. They can be formed using nozzles with orifice diameters that are large enough to accommodate protein microcrystals, and we successfully jetted C-phycocyanin microcrystals in preparation for future tests at a beamline. It is expected that the low scattering background produced by these thin jets will increase the signal-to-noise ratios of measured Bragg spots, to further improve diffraction measurements of sub-micrometer sized protein crystals~\cite{Williamson:2023}. By using viscoelastic jets to deliver \blue{viscous slurries of nanocrystals of high number concentration, we expect to significantly improve the jet stability and therefore the overlap of the sample stream with the X-rays, which will lead to competitive ``hit rates" at which diffraction data is collected.} This will further enable high-accuracy investigations of, for instance, subtle photoinduced density changes. X-ray hit rate statistics for different flow rate ratios will be evaluated and discussed in a forthcoming article.    

\begin{acknowledgements}
The authors thank Julia Maracke (Center for Free-Electron Laser Science, DESY) and Agnieszka Wrona (European XFEL) for technical support in device fabrication. Sample preparation was conducted in the XFEL Biology Infrastructure (XBI) BioLab of the European XFEL and we thank \blue{Christina Schmidt for support in sample characterization}.
\end{acknowledgements}

\begin{funding}
This work was financially supported by the Spanish Ministry of Science, Innovation and Universities (grant no. PID2022-140951OB/AEI/10.13039/501100011033/FEDER, UE). We acknowledge additional support by DESY (Hamburg, Germany), a member of the Helmholtz Association HGF and by the Cluster of Excellence ‘Advanced Imaging of Matter’ of the Deutsche Forschungsgemeinschaft (DFG)—EXC 2056—project ID 390715994.
\end{funding}

\ConflictsOfInterest{There are no conflicts of interest.}

\DataAvailability{Data supporting the results can be obtained from the authors upon reasonable request.}

%\bibliography{intro,central} % basename of .bib file

\end{document}